  \documentclass[sigconf]{acmart}
\usepackage{tcolorbox}
\usepackage{makecell} 
\usepackage{graphicx}
\usepackage{textcomp}
\usepackage{xcolor}
\usepackage{listings}
\usepackage{pifont}
\usepackage{booktabs}
\usepackage{multirow}
\usepackage{pifont} % For tick and cross marks
\usepackage{setspace}
\usepackage{mdframed}
\AtBeginDocument{%
  }

\setcopyright{acmlicensed}
\copyrightyear{2025}
\acmYear{2025}
\acmDOI{XXXXXXX.XXXXXXX}
\acmConference[EASE 2025]{The 29th International Conference on Evaluation and Assessment in Software Engineering}{17–20 June, 2025}{Istanbul, Türkiye}
\acmISBN{978-1-4503-XXXX-X/2018/06}

\begin{document}

\title{Can We Enhance Bug Report Quality Using LLMs?: An Empirical Study of LLM-Based Bug Report Generation}
\author{Jagrit Acharya}
\orcid{0009-0008-0302-6130}
\affiliation{%
  \institution{University of Calgary}
  \city{Calgary}
  \country{Canada}
}
\email{jagrit.acharya1@ucalgary.ca}

\author{Gouri Ginde}
\orcid{0000-0001-7519-3503}
\affiliation{%
  \institution{University of Calgary}
  \city{Calgary}
  \country{Canada}
}
\email{gouri.deshpande@ucalgary.ca}


\begin{abstract}
Bug reports contain the information developers need to triage and fix software bugs. However, unclear, incomplete, or ambiguous information may lead to delays and excessive manual effort spent on bug triage and resolution. In this paper, we explore whether Instruction fine-tuned Large Language Models (LLMs) can automatically transform casual, unstructured bug reports into high-quality, structured bug reports adhering to a standard template. We evaluate three open-source instruction-tuned LLMs (\emph{Qwen 2.5, Mistral, and Llama 3.2}) against ChatGPT-4o, measuring performance on established metrics such as CTQRS, ROUGE, METEOR, and SBERT. Our experiments show that fine-tuned Qwen 2.5 achieves a CTQRS score of \textbf{77\%}, outperforming both fine-tuned Mistral (\textbf{71\%}), Llama 3.2 (\textbf{63\%}) and ChatGPT in 3-shot learning (\textbf{75\%}). Further analysis reveals that Llama 3.2 shows higher accuracy of detecting missing fields particularly Expected Behavior and Actual Behavior, while Qwen 2.5 demonstrates superior performance in capturing Steps-to-Reproduce, with an F1 score of 76\%. Additional testing of the models on other popular projects (e.g., Eclipse, GCC) demonstrates that our approach generalizes well, achieving up to \textbf{70\%} CTQRS in unseen projects' bug reports. These findings highlight the potential of instruction fine-tuning in automating structured bug report generation, reducing manual effort for developers and streamlining the software maintenance process.

\end{abstract}




\maketitle
Can We Enhance Bug Report Quality Using
LLMs?: An Empirical Study of LLM-Based Bug
Report Generation

\section{Introduction}

Bug reports are essential in software maintenance, providing developers with critical information to identify, triage, and resolve software defects \cite{good_report}. A bug report is a record of a software fault or defect that is created by an end-user or a tester \cite{ahmed2014impact}. However, the effectiveness of bug reports is often hindered by ambiguity, incompleteness, or inconsistency in the information provided by reporters \cite{aranda2009secret,ko2010power}.
Well-structured reports that clearly articulate observed behavior (OB), expected behavior (EB), and steps to reproduce (S2Rs) minimize ambiguity and enable developers to resolve issues without much discussion \& clarification \cite{wang2015}.

% One possible reason for this issue is that reporters have different levels of experience and understanding of the software, making it difficult for them to communicate the necessary details to developers \cite{vyas2014bug,moran2015auto}. Additionally, the most important information for resolving a bug, such as steps to reproduce, actual results, and test cases, is often challenging for reporters to provide \cite{erfani2014works,chen2014ar,joorabchi2013real}. Research indicates that inadequate tool support during the bug reporting process significantly contributes to inaccuracies in bug reports \cite{erfani2014works,just2008towards}.

Challenges in bug reporting persist due to factors such as varying reporter experience and difficulty in providing essential details like reproduction steps and expected outcomes \cite{vyas2014bug,moran2015auto,chen2014ar,joorabchi2013real}. Adding to this problem, the lack of tool support during report creation further undermines accuracy \cite{erfani2014works,just2008towards}. To address these challenges researchers have explored improving bug report quality by detecting weak descriptions \cite{schuegerl2008enriching}. Some studies check if a report includes key details like observed behavior, expected behavior, and steps to reproduce the issue \cite{chaparro2017detecting, chaparro2019assessing} using Natural Language Processing (NLP) based approaches. Others provide an overall quality assessment of the bug report and offer general suggestions for improvement \cite{schuegerl2008enriching, good_report}. Advancements in NLP have led to the development of large language models (LLMs), which are transformer-based neural networks capable of predicting the next token based on the preceding context \cite{sarkar2022like}. These models comprehend context and execute assigned tasks through prompts. A similar architecture has been employed by Bo et al. \cite{Chatbr} to generate missing information in bug reports using ChatGPT. However, ChatGPT has been found to generate incorrect information \cite{chatgpt_hal, chatgpt_hal2} and faces limitations due to data privacy concerns \cite{Cai2024FCodeLLMAF} in Software Engineering.

% There are numerous instances where steps to reproduce (S2Rs) are unclear, incomplete, or ambiguous, preventing developers from effectively replicating issues and resolving software bugs \cite{fazzini2018automatically,good_report}. Additionally, important information is often missing in incomplete bug reports, leading to non-reproducible \cite{erfani2014works} and unresolved bugs \cite{zimmermann2012characterizing}. Our fine-tuned model addresses this problem by guiding reporters on missing information, ensuring more comprehensive bug reports.

% \textbf{Hence, in this study, our aim is} to provide an approach to transforming unstructured bug reports into structured bug reports according to standardized template formats while also highlighting missing pieces of information to the reporter before submission of the bug report using open source large language models. 

\textbf{Hence, in this study} our aim is to provide an approach to transforming unstructured bug reports into structured bug reports according to standardized template formats while also highlighting missing pieces of information to the reporter before submission of the bug report using open source large language models locally.

\noindent \textbf{Our research contributions are as follows:} 
\vspace*{-2pt} 
\begin{itemize}
% \vspace{5pt}
% \\.
\vspace*{-2pt} 
\item We provide empirical evidence to show that the instruction fine-tuned LLMs perform close to state-of-the-art (SOTA) model, ChatGPT-4o (hereafter referred to as ChatGPT) in generating high-quality bug reports based on measures such as SBERT, ROUGE-1, and CTQRS scores.

\item To the best of our knowledge, we are the first to demonstrate how LLM models can transform the reporter (natural-language-based) summary into a structured bug report as per the bug template format. This contribution is particularly significant as bug report quality directly impacts all other research domains, such as bug triage, assignments, duplication detection and prioritization.

\item We show evidence for the effectiveness of Cross Platform learning (LLMs trained on bug reports from larger projects are used to generate bug reports for smaller projects). As such, our results show that instruction fine-tuned models can generalize well and perform significantly better for projects without training data.

\item As a contribution to open science, we make our complete dataset and source code public for researchers to replicate our study and utilize the dataset for other explorations.\footnote{\url{https://github.com/GindeLab/Ease_2025_AI_model}}

\end{itemize}

 % Apart from a lack of awareness about good bug writing guidelines, the knowledge gap between end-users and developers also plays a significant role in the quality of bug reports. Many end users lack familiarity with the internals of the system, which could lead to reports that do not contain the necessary information developers need to diagnose and fix bugs \cite{Song2022}.
% todo refine this upper wala example daal sakte ho non experience wala
 
% No location found to fit this //todo

 % In such scenarios, it was observed that developers tend to consider stack traces and steps to reproduce a bug to help resolve them, which is a time and effort-consuming endeavour \cite{breu2010information}, \cite{zimmermann2012characterizing}. Thus, well-structured bug reports that adhere to bug writing guidelines are essential for effective bug triage and resolution. 

% // todo Time wale examples ki baat karo yaha
 % Need of structure report
% A well-structured bug report is crucial for effective software maintenance tasks such as bug triaging and fixing. Paikari \cite{Paikari2023} emphasizes the significance of structured bug reports in determining their actionability. Our fine-tuned model generates structured bug reports from casual written bug reports.
%  % Todo Vo bhi likho ki soft maintiance task ashe se ho jate hai 

 \section{Need for this Study}

This section presents a motivating example to illustrate our approach and compares it with existing methods for improving bug report quality. We investigate how instruction fine-tuned LLMs perform against state-of-the-art models like ChatGPT in generating structured bug reports, their effectiveness in generalizing across different software projects, and their ability to identify missing information while mapping summaries to structured components. Our evaluation, based on instruction fine-tuning, uses both qualitative and quantitative metrics, as shown in Table \ref{tab:comparison}.

% In this section, we present a motivating example to illustrate our approach and compare it with other existing methods used for improving bug report quality.
% In this paper, we investigate: (1) how instruction fine-tuned
% LLMs perform compared to a state-of-the-art model like ChatGPT4o (hereafter referred to as ChatGPT) in generating structured bug
% reports, (2) the effectiveness of these models in generalizing across
% different software projects, and (3) their ability to identify missing
% information and map summaries to structured report components.

% We achieve this through instruction fine-tuning and evaluate our models using both qualitative and quantitative metrics as shown in Table \ref{tab:comparison} .

\begin{figure}[h!]
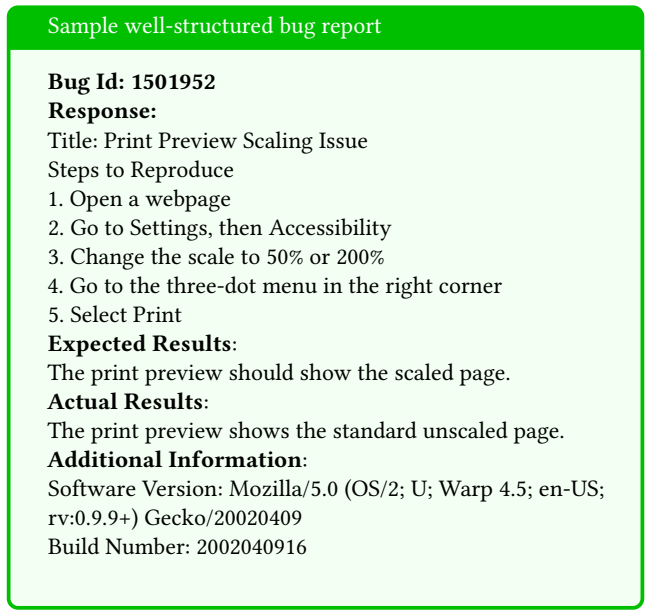

    \centering
    \begin{tcolorbox}[colback=green!5!white, colframe=green!75!black, title=Sample well-structured bug report]
    \textbf{Bug Id: 1501952} \\
    \textbf{Response:} \\
Title: Print Preview Scaling Issue\\
    Steps to Reproduce \\
    1. Open a webpage \\
    2. Go to Settings, then Accessibility \\
    3. Change the scale to 50\% or 200\% \\
    4. Go to the three-dot menu in the right corner \\
    5. Select Print \\
    \textbf{Expected Results}: \\
    The print preview should show the scaled page. \\
    \textbf{Actual Results}: \\
    The print preview shows the standard unscaled page. \\ 
    \textbf{Additional Information}: \\
    Software Version: Mozilla/5.0 (OS/2; U; Warp 4.5; en-US; rv:0.9.9+) Gecko/20020409 \\
    Build Number: 2002040916 \\
    \end{tcolorbox}
    \caption{This is an example of a bug report generated from a fine-tuned
Mistral 7B model, based on the unstructured report}
    \label{green_output}
       \vspace{-3mm}
\end{figure}

\noindent \textbf{Motivating example:}
Figure \ref{red_input} presents a sample bug report from Bugzilla, which lacks explicitly stated steps to reproduce the issue. As a result, the developer had to request clarification, leading to a delay in resolving the bug, which was ultimately fixed only after the reporter provided clear reproduction steps two months later. In contrast, Figure \ref{green_output} illustrates a well-structured bug report, similar report was automatically triaged and resolved within five days, requiring minimal discussion or clarification with the reporter.
\begin{figure}[h!]
    \centering
    \begin{tcolorbox}[colback=red!5!white, colframe=red!75!black, title=Sample unstructured (lacking standard format) bug report]
    \textbf{Bug Id: 1805934} 

User Agent: Mozilla/5.0 (X11; Linux x86\_64; rv:108.0) Gecko/20100101 Firefox/108.0

Steps to reproduce:

For a few months now, I've been suffering an intermittent problem: every now and again, all drop-down controls in Firefox would break. Menus would no longer work, drop-down selects on web pages would fail, extension menus would fail, and the hamburger menu would fail. The visible behavior is that the drop-down is drawn but then immediately erased as if I had clicked elsewhere in the window. The only fix for the problem is to restart Firefox.

Recently, I realized something: every time I restarted to fix the issue, Firefox would bring up the dialog saying it was installing the latest update. And I never get a dialog to tell me that an update is available.

So what seems to be happening is: every time Firefox detects an available update, something breaks and all menus and drop-downs stop working.

Today was even worse: restarting didn't show the updating dialog, and as soon as I went to any web page, all the drop-downs broke again. So I wondered if I was wrong about the cause... but I cleared all local data (cache, cookies, the lot) and restarted one more time -- and suddenly I got the updating dialog, and now drop-downs work again.

Obviously this is absolutely infuriating. I'd like to do anything I can to help you track down and fix the problem.
   \end{tcolorbox}
\caption{\textbf{This is an example of a low-quality bug report, as it does not follow the defined Bugzilla bug report template.}}

    \label{red_input}
    \vspace{-3mm}
\end{figure}

\begin{table}[h]
    \centering
    \renewcommand{\arraystretch}{1.0} % Adjust row height slightly, reduced from 1.1
    \setlength{\tabcolsep}{5pt} % Adjust column spacing
    \vspace{-5pt} % Reduce space above the table
    \begin{tabular}{|l|c|c|c|}
        \hline
        \textbf{Papers} & \textbf{Model} & \textbf{Metrics} & \textbf{Adaptation} \\
        \hline
        GIRT\cite{girt-model}   & Open Source & Quantitative & Fine-tuning \\
        \hline
        ChatBR \cite{Chatbr} & ChatGPT & Quantitative & Few-Shot \\
        \hline
        BugBlitz \cite{bugblitzaiintelligentqaassistant} & Open Source & Quantitative & Fine-tuning \\
        \hline
        \textbf{Our Study} & \textbf{Open Source} & \makecell{\textbf{Qualitative \&} \\ \textbf{Quantitative}} & \textbf{Fine-tuning} \\
        \hline
    \end{tabular}
\caption{Comparison of prior work with our approach in terms of model type, evaluation metrics employed, and adaptation techniques applied.}

    \label{tab:comparison}
    \vspace{-25pt} % Reduce space below the table
\end{table}

\section{Preliminaries} \label{Pre}
In this section, we describe the instruction fine-tuning of LLMs, the specific LLM used in our study, and the evaluation metrics CTQRS and related concepts.

 \textbf{Instruction fine-tuning: } 
 % \textbf{EXPLAIN what is it and how it is different from generic fine-tuning}
 Instruction fine-tuning trains a language model to follow specific instructions by learning from examples \cite{ins_tune1}. For instance, we provide the model with pairs of bug report summaries and well-written bug reports. By learning from these examples, the model can automatically convert any new unstructured report from the reporter into a complete, well-structured bug report following a bug report template. This differs from generic fine-tuning, which simply adapts the model to specific data without explicitly teaching it to execute instructions; instruction fine-tuning makes the model better at understanding and performing the desired transformation as per the given guidelines \cite{köksal2024longformeffectiveinstructiontuning}. After fine-tuning with annotated datasets containing instructional data, language models show an enhanced ability to follow general language instructions \cite{honovich2022unnaturalinstructionstuninglanguage}. This method, known as instruction-tuning, enhances the controllability of LLMs via natural language commands, thus significantly improving their performance and ability to generalize across unseen tasks 
 \cite{IT4,wang2022supernaturalinstructionsgeneralizationdeclarativeinstructions}.

\textbf{LLM Models:} We conducted supervised fine-tuning using Low-Rank Adaptation (LoRA) \cite{LORA} on three top-performing open-source instruction-tuned language model: \url{unsloth/Mistral-7B-Instruct-v0.3} \cite{Mistral_7b_instruct_v0_3}, \url{unsloth/Qwen2.5-7B-Instruct} \cite{qwen2.5}, and \url{unsloth/Llama-3.2-3B-Instruct} \cite{unsloth_llama_3_2_3b_instruct}.
These models were selected based on their strong ranking on the Hugging Face Open LLM Leaderboard \cite{beeching2023open} as of February 2025 and widespread use in recent studies \cite{jindal2024birbalefficient7binstructmodel, hawkins2024effectfinetuninglanguagemodel, Mistral_exp1}. They also vary in size (7B and 3B parameters), allowing us to study how scale affects performance. We used the Unsloth training framework \cite{Unsloth_GitHub} for efficient fine-tuning, as it significantly reduces VRAM usage and speeds up training. For comparison, we also used Llama 3 \cite{dubey2024llama3herdmodels} for generating unstructured bug reports and ChatGPT \cite{openai2023ChatGPT} for few-shot learning tasks.

\textbf{CTQRS:} CTQRS (Crowdsourced Test Report Quality Score), developed by Zhang et al. \cite{CTQRS}, is a bug-report quality assessment framework that systematically scores bug reports by combining morphological, relational, and analytical indicators through dependency parsing. We re-implemented all the 13 rules defined by the authors to determine the score of the bug reports using python.

\begin{figure*}[!htpb]
\centering
\includegraphics[scale=.15]{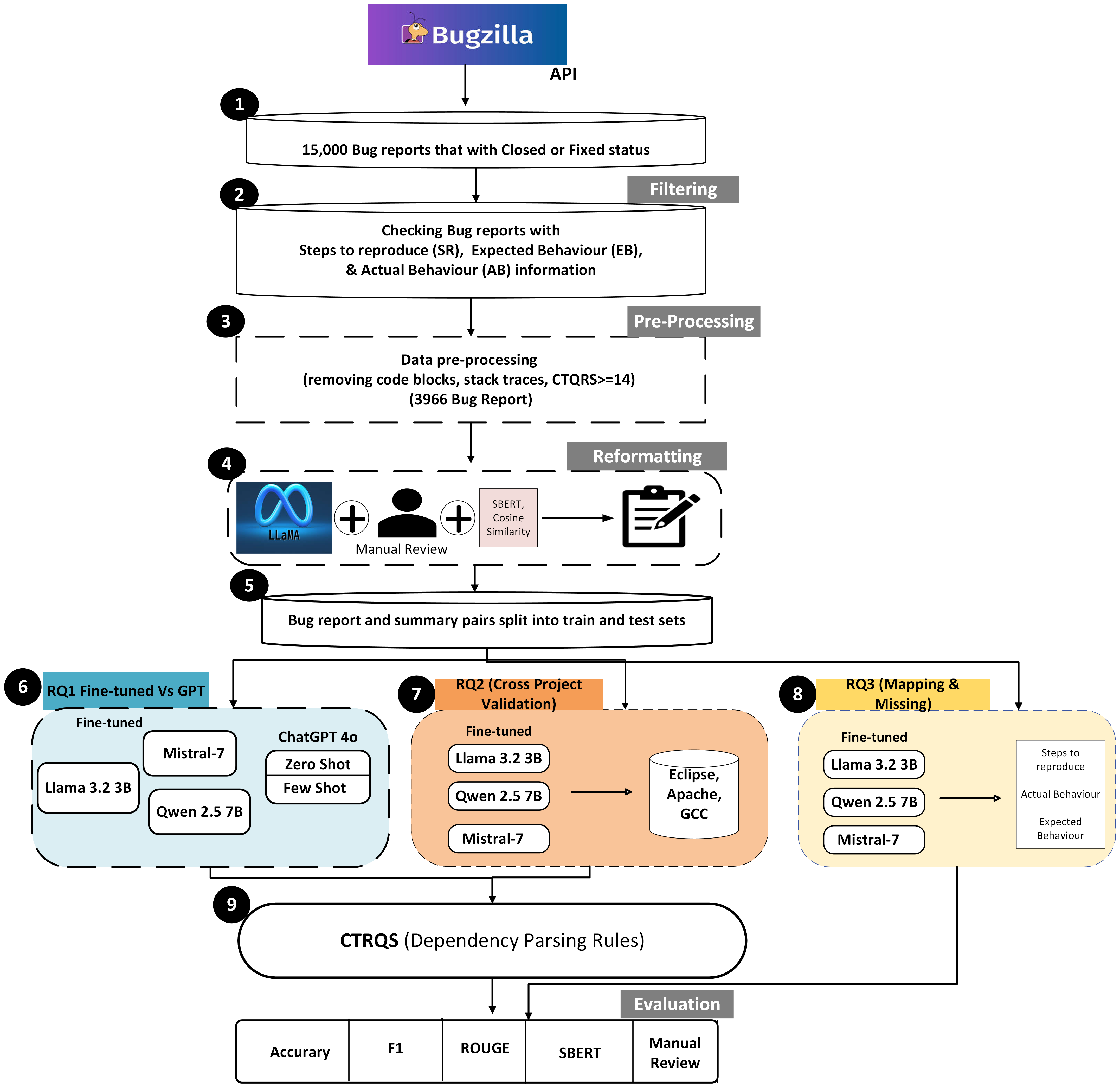}
\caption{Architecture for Generating High-Quality Bug Reports from Unstructured Bug Reports Using Fine-Tuned Large Language Models}
\label{Arch_diag}
\end{figure*}

\section{Research Questions} \label{RQ}
We aims to evaluate the efficacy of instruction-fine-tuned large language models (LLMs) in generating structured, high-quality bug reports compared to general-purpose models like ChatGPT, assess their cross-project generalizability when applied to diverse open source software projects, and analyze their ability to identify missing information from user-provided bug reports while accurately mapping details to standardized bug report components.
\begin{enumerate} 
%     \item [\textbf{RQ1:}] \textbf{Does instruction fine-tuning improve the quality of LLMs generated bug reports? }
%     % compared to those from pre-trained (base) hs regarding BLEU, ROUGE-1 and METEOR scores?
% \\
%     \textbf{Rationale:} Instruction fine-tuning helps the models generate and adapt to the domain-specific knowledge and jargon as well as output in a specific format. Thus, for this research question, we explore its utility to automatically generate bug reports adhering to a pre-defined report template.

% RQ: What is the effectiveness of the fine-tuned LLM in mapping summary information to structured bug reports and in inferring expected results from actual results?

% Rationale: Evaluating the LLM’s mapping accuracy ensures the generated reports are detailed and useful, while testing its inference capability assesses its potential to assist in scenarios where bug reports are partially documented.

    \item [\textbf{RQ1:}]\textbf{How do fine-tuned LLM models perform compared to the GPT model in generating bug reports? }
    % according to the bug report template regarding BLEU, ROUGE-1 and METEOR scores?
    \\
    \textbf{Rationale:} ChatGPT models which are trained on a large corpus need little to no prior knowledge of the task in hand. Thus, through this RQ, we evaluate the effectiveness of instruction fine-tuned models against a widely used GPT model in SE research. This comparison helps determine if specialized fine-tuning provides any significant advantages over general-purpose models in generating structured, high-quality bug reports.

    \item [\textbf{RQ2:}] \textbf{ How effective is cross-project prediction in generating structured bug reports across different software projects? } \\
    \textbf{Rationale:} Cross-project prediction enables models trained on bug reports from one project to be applied to different projects, assessing their generalizability. Evaluating a fine-tuned model trained on larger projects will help determine if learned patterns can effectively transfer across different software projects.
      \item [\textbf{RQ3:}] \textbf{ How effective is the fine-tuned LLM in identifying missing information from summaries and mapping unstructured report information to structured bug report components?} \\
    \textbf{Rationale:} Evaluating the fine-tuned LLM’s effectiveness in identifying missing information mapping accuracy ensures the generated reports capture all details from the unstructured report, correctly identifies the missing information and maps them onto the respective component of bug report. It helps identify the bug report components where the LLM demonstrates the strongest and weakest effectiveness in mapping and detecting missing information.
%  \item [\textbf{RQ4:}] \textbf{ How do different prompts affect the
% generation results?} \\
%     \textbf{Rationale:} We examine the influence of prompt design on LLMs generated output. Beginning with a basic template, we iteratively refine it by framing LLMs as senior software engineers with expertise in bug report writing. Additionally, we integrate specific requirements for OB, EB, and S2R and Additional Information aligning with ChatGPT's best practices \cite{ChatGPT_PromptEng_2023}.
   
\end{enumerate}

\section{Methodology} \label{Methodology}
% In this section we elaborate on the dataset and our study design in detail. 
In this section, we discuss the dataset, preprocessing steps, data generation, prompt design, model fine-tuning, implementation details, and evaluation metrics used in our study  (as shown in Figure \ref{Arch_diag}).

\noindent\textbf{Dataset and Pre-Processing:}\label{dataset}
 We mined a dataset comprising the recent 15,000 fixed bug reports from Bugzilla, an online bug tracking system, that were ``fixed" and ``closed", (as considered in the previous work \cite{closed_bugs_gather}). The dataset was gathered using Bugzilla API over multiple iterations. First, relevant bug reports' metadata was gathered utilizing the ``Get All Data" API call. Then, using the ``Get All Comments" APIs, all the details regarding bug reports were gathered (Step \ding{202}). The dataset includes fields such as Bug ID, Comment ID, Comment, Priority, Severity, Status etc. Our primary focus was the Comment field, as it contained the key bug report details required for fine-tuning.

% \begin{table}[ht] % or [htbp] for more flexibility
% \renewcommand{\arraystretch}{1.3}
% \centering
%  \caption{Detailed properties of the dataset, including descriptions for each field extracted from Bugzilla and used in our study.}
% \begin{tabular}{p{1.5cm} p{4.5cm} p{1cm}}
% \textbf{Property} & \textbf{Description}                                 & \textbf{Data type} \\ \hline
% \textit{Bug ID}          & Unique identifier of a bug report                                  & Number                                  
%  \\ \hline
%  \textit{Comment ID}          & Unique identifier of a comment                                    & Number                                  
%  \\ \hline
% Summary                   & A brief title of the bug report                     & String                                         \\ \hline
% Product                   & The product or project the bug/issue is associated with & String  \\ \hline
% Comment (Bug Report)      & Comment is a discussion related to a bug; the bug reports are mentioned in the first comment    & String    \\ \hline

% % \hline
% % Llama3 Summary            & Summary generated using Llama 3 from the Bug report, used as an input for fine-tuning  & String
% \end{tabular}
% \label{Table:Datacontent}
% \end{table} 
% Table \ref{Table:Datacontent} outlines the properties included in our dataset, along with their descriptions and data types. These properties encompass unique identifiers for bugs and comments (bug ID and comment ID), textual summaries, associated products, detailed comments, completion times.
Not all bug reports contained the necessary information outlined in the Bugzilla bug report guidelines \cite{Bugzilla_guidelines}, which advises reporters to include steps to reproduce (S2Rs), actual results (AR), expected results (ER), and any additional relevant information in their bug report. Thus, to curate the high-quality training dataset,firstly, we employed regular expressions to filter the bugs whose bug reports had descriptions, S2Rs, EB, AB, and additional information, as shown in (Step \ding{203}). Secondly, bug reports containing stack traces or code snippets were similarly excluded. This decision was motivated by the potential for these elements to introduce noise and complexity \cite{CTQRS}, thereby negatively impacting the fine-tuning process of our model. Afterwards filtered bug reports with a CTQRS score greater than 14 as they are considered good by Zheng et al. \cite{CTQRS} (Step \ding{204}).
After these filtrations, we had 3,966 Bugs with all the required information (Step \ding{203}), out of which 200 reports were manually reviewed to check if they were of the desired quality.

% Bugzilla consists of bug reports of all the products offered by Mozilla. The product distribution of the 10,000 bug reports we gathered is shown in Figure \ref{fig:10K_PDF}, where Core is the largest project, followed by Firefox. Product distribution of our curated, well-structured bug reports data, also referred to as training data, is as shown in Figure  \ref{fig:3K_PDF}. In this distribution, Firefox has the most well-structured bug reports, followed by the Core project.
\begin{lstlisting}[
    float,
    frame=single,
    breaklines=true,
    caption={Llama 3 prompt example for generating unstructured bug report},
    label={llama_prompt}
]
Please rewrite the following bug report in a natural, conversational tone, 
as if you're explaining it to someone casually. Keep the essence of the report 
intact, but restructure it in a way that sounds like something an average 
person would write, while still using the original wording from the report as 
much as possible. Focus on maintaining the original details and key points 
without changing much. Provide only the one rewritten paragraph with everything, 
no additional explanation. 

Bug report: {text}
\end{lstlisting}

\noindent \textbf{Synthetic (pseudo-ground truth) data generation:}
  Recent studies have successfully utilized LLMs to generate factually consistent summaries \cite{Tang2023DoesSD} and generate data in the domain of healthcare \cite{Gekhman2023TrueTeacherLF}. Taking inspiration from these studies, we first conducted multiple experiments with different keywords and we got best results when we requested with please keyword and final evaluations to design the prompt as shown in Figure \ref{llama_prompt} (to generate an unstructured from a well-structured bug report). Further, utilizing this prompt with the state-of-the-art Llama3 model \cite{dubey2024llama3herdmodels}, we generated summaries for all 3,966 well-structured bug reports in our training set. To ensure the generated reports were closely aligned with the original reports, we manually verified 200 reports computed SBERT \cite{SBERT} and cosine similarity \cite{cosine} scores as shown in step \ding{205} of Figure \ref{Arch_diag}. Each unstructured report was generated three times, and only those with an SBERT similarity exceeding 85\% and a cosine similarity above 80\% were retained. The final dataset comprised 3,903 well-structured bug reports paired with their summaries, serving as synthetic pseudo-ground truth for instruction fine-tuning tasks.
%   \begin{figure}[h!]
%     \centering
%     \begin{mdframed}[linewidth=0.5pt, innermargin=5pt, innerleftmargin=5pt, innerrightmargin=5pt, backgroundcolor=white, linecolor=black]
%     \textbf{Llama 3 Prompt Example} % Simple heading inside the box
%     \vspace{5pt} % Adds a little space before the text
%     \small
%     \begin{lstlisting}[breaklines=true]
% Please rewrite the following bug report in a natural, conversational tone, 
% as if you're explaining it to someone casually. Keep the essence of the report 
% intact, but restructure it in a way that sounds like something an average 
% person would write, while still using the original wording from the report as 
% much as possible. Focus on maintaining the original details and key points 
% without changing much. Provide only the one rewritten paragraph with everything, 
% no additional explanation. 

% Bug report: {text}
%     \end{lstlisting}
%     \end{mdframed}
%     \caption{Llama 3 Prompt Example for generating Bug Report summaries in Everyday Language.}
%     \label{llamaprompt}
% \end{figure}

\begin{lstlisting}[
    float,
    frame=single,
    breaklines=true,
    caption={Alpaca prompt template used to fine-tune open-source LLMs.},
    label={lst:bug-report-prompt}
]
alpaca_prompt  You are a senior software engineer specialized in generating detailed bug reports.
### Instruction:
Please create a bug report that includes the following sections:
1. Steps to Reproduce (S2R): Detailed steps to replicate the issue.
2. Expected Result (ER): What you expected to happen.
3. Actual Result (AR): What actually happened.
4. Additional Information: Include relevant details such as software version, build number, environment, etc.

If any of these sections are missing from the provided report, explicitly notify the user which information is missing.

### Input:
{unstructured_report}

### Response:
{Bug_report}
    \end{lstlisting}
    % \end{mdframed}

\noindent\textbf{Data Splitting:}\label{PP} Step \ding{206} in Figure \ref{Arch_diag} illustrates the data splitting process, where the data was randomized and then split into training, testing and validation sets, where training was 80\% of the data, comprising 3,122 rows, testing 10\% with 391 rows, and the remaining 10\% was validation with 390 rows. We finetuned our model using 4 cross-validation. \\
% \vspace{-2mm}
\textbf{Prompt Design:} \label{prompt}
A prompt \cite{liu2023pre} serves as a set of instructions that directs LLMs to generate a specific desired output \cite{bang2023multitask}. The effectiveness of an LLM's performance on the same task can vary depending on the prompt used \cite{kojima2022large}, making it essential to craft precise prompts. In our approach, we employ a single-round dialogue interaction to formulate prompts, utilizing the standard Alpaca-LoRA template \cite{Tloen} as shown in Figure \ref{lst:bug-report-prompt}. Additionally, we adopt a strategy similar to that used by Bo et al. \cite{Chatbr} to create an effective prompt template for fine-tuning:(1) providing important task-related context as much as possible; (2) assigning LLMs a specified role for our task (Senior Software Engineer); (3) using separators in the prompt to indicate different parts of the input; (4) formatting the LLM’s output in a standardized JSON structure for better analysis.; and (5) ensuring the prompt is both concise and accurate to fit within the LLM's input token limitations. The prompts have been structured using the standard Alpaca-LoRA template \cite{Tloen} meticulously encoded through the model's tokenizer.This includes adding the \texttt{<|begin\_of\_text|>} token (equivalent to the BOS token) and the \texttt{<|eot\_id|>} token (which signifies the end of the message in turn).
All the parameters used for this step are reported in the example of fine-tuning using Unsloth and the TRL SFTTrainer \cite{Supervised_Fine_tuning_Trainer} available on our GitHub repository.\footnote{\url{https://github.com/GindeLab/Ease_2025_AI_model}}

\noindent \textbf{Instruction fine-tuning:}
We used  Parameter Efficient Fine-Tuning (PEFT), which doesn't fine-tune the entire model but modifies several parameters to adapt the models for different applications \cite{mangrulkar2022peft}. This approach helps reduce the substantial expenses linked to full-fine-tuning and ensures that fine-tuning is feasible even with constrained storage and processing power. Low-rank adaptation (LoRA) \cite{LORA}, a prominent PEFT technique, reduces trainable parameters by incorporating low-rank trainable matrices within the attention layers of the Transformer model and freezing the model's weights.

We have fine-tuned the top three widely used models in the literature and Huggingface leaderboard: Qwen 2.5-7B, Mistral-7B, and Llama 3.2B \cite{beeching2023open,Mistral_exp1,jindal2024birbalefficient7binstructmodel,hawkins2024effectfinetuninglanguagemodel}. To improve efficiency and reduce resource consumption, we adopt the Unsloth framework to optimize the Low-Rank Adaptation (LoRA) method for fine-tuning the models, setting the rank to 16. We specifically target LoRA modules such as \texttt{'q\_proj'}, \texttt{'k\_proj'}, \texttt{'o\_proj'}, \texttt{'v\_proj'}, \texttt{'down\_proj'}, \texttt{'gate\_proj'}, and \texttt{'up\_proj'}.

Following standard fine-tuning hyperparameters, we train the Mistral-7B and Qwen-2.5 B model for 3 epochs with a learning rate of 2e-4 and a batch size of 8 examples. We also fine-tuned the Llama-3.2 3B models in the subsequent analysis. For these models, we conducted training for 3 epochs with a learning rate of 3e-3 and a batch size of 8 samples.

The number of learning rate, LORA rank and epochs were determined based on experimental observations; we noticed that the model's performance on the validation set plateaued after 3 epochs, indicating that additional training did not yield significant improvements and a 4-Cross validation was applied.\\

\noindent\textbf{Implementation Details}\label{Implement}
The hardware configuration used for fine-tuning was RTX 4090 GPU, with 32 GB of RAM. The models were fine-tuned using the Unsloth framework \cite{Unsloth_GitHub}.

\noindent \textbf{Evaluation metrics:}
As illustrated in \ding{210}, all models were evaluated using established metrics widely employed in similar studies \cite{score1,score2,score3,score5}: specifically, ROUGE-1 \cite{lin-2004-ROUGE}, SBERT \cite{SBERT} and Cosine Similarity \cite{cosine}.  along with Accuracy and F1 Score.

Although ROUGE (Recall-Oriented Understudy for Gisting Evaluation) is a widely used metric for summarization tasks, it is applicable to evaluating paraphrases. ROUGE-1 measures recall through matched unigrams, and we employ this variant in our assessments.

We implemented the bug report quality metrics score proposed by Zhang et al. \cite{CTQRS}, which evaluates reports based on Atomicity, Conciseness, Completeness, Understandability, and Reproducibility using dependency parsing.

% METEOR (Metric for Evaluation of Translation with Explicit ORdering) uses a weighted F-score that includes unigram mapping and penalizes word order errors.

For RQ1 \ding{207}, we compared the quality of bug reports generated by fine-tuned models with those from ChatGPT. For RQ2 \ding{208},we assessed the generalization capability of the fine-tuned models. For RQ3\ding{209},  we evaluated the accuracy of the fine-tuned models for mappings of unstructured bug reports to high-quality bug reports and testing model's missing information identification. 

\vspace*{-2pt} 
% % Document content
% \begin{figure}[h!]
%     \centering
%     \begin{tcolorbox}[colback=gray!10, colframe=gray!80, title=Fine Tuning Prompt Example]
%     \begin{lstlisting}[breaklines=true]

% alpaca_prompt  You are a senior software engineer specialized in generating detailed bug reports.
% ### Instruction:
% Please create a bug report that includes the following sections:
% 1. Steps to Reproduce (S2R): Detailed steps to replicate the issue.
% 2. Expected Result (ER): What you expected to happen.
% 3. Actual Result (AR): What actually happened.
% 4. Additional Information: Include relevant details such as software version, build number, environment, etc.

% If any of these sections are missing from the provided summary, explicitly notify the user which information is missing.

% ### Input:
% {Summary}

% ### Response:
% {Bug_report}

%     \end{lstlisting}
%     \end{tcolorbox}
    
%     \textbf{\caption{Instruction Fine-Tuning Prompt for Comprehensive Bug Report Generation. Structured to include reproduction steps, expected and actual results, and additional relevant information while ensuring completeness.}}
%     \label{fine_tuning_prompt}
%     \vspace{-5mm}
% \end{figure}
 % \vspace{-6mm}
\section{Results} \label{results}
\begin{figure*}[!t]
\begin{minipage}{.45\textwidth}
    \centering
\includegraphics[scale=.3] {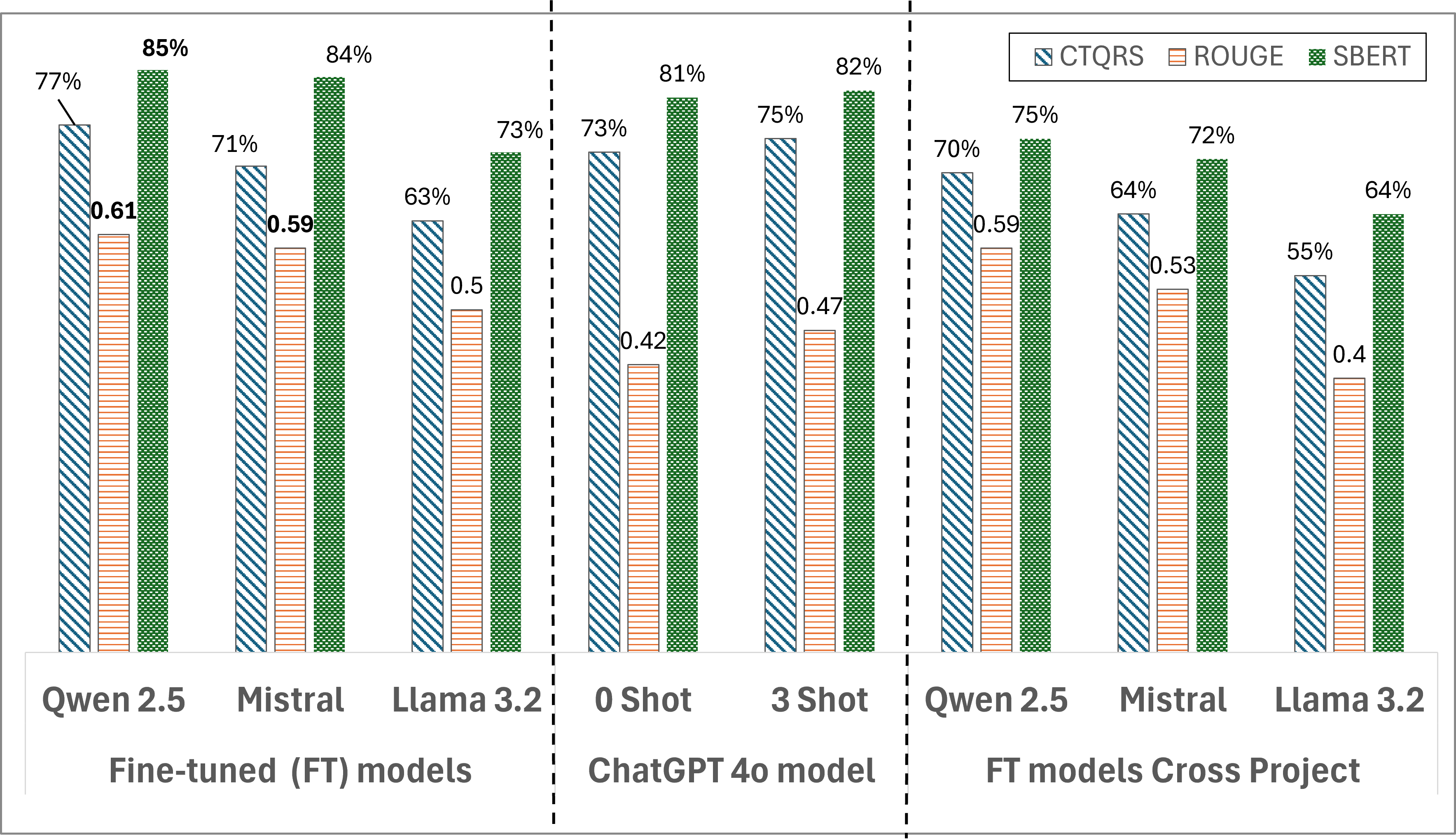}
   \caption{RQ1 and RQ2: Comparing the performance of fine-tuned models with base models and ChatGPT 4o on test dataset}
    \label{fig:RQ1nRQ2}    
    \end{minipage}
    \hfill
    \begin{minipage}{.45\textwidth}
    \centering
 \includegraphics[scale=.45, trim=1.5cm 0 0 0]{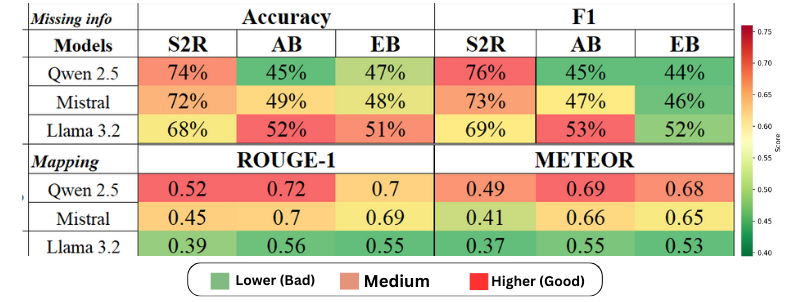}
\caption{RQ3 – Heat-map. Upper part (“Missing info”): shows how accurately the model can flag missing fields (higher = better). Bottom part (“Mapping”): shows how well the model maps content from user text to structured report fields (higher = better).}

    \label{fig:RQ3}
    \end{minipage}
    \vspace{-3mm}
\end{figure*}
We evaluated our fine-tuned models using CTQRS, ROUGE-1, and SBERT by passing unstructured test reports through the model to generate structured reports. The CTQRS score measures the quality of a bug report out of 17, based on the rules discussed in Section \ref{Pre}. As shown in Figure \ref{fig:RQ1nRQ2}, the Qwen 2.5 model achieved an average score of approximately 77\% on the test dataset, meaning the generated reports received an average of 13 out of 17 points. The ROUGE score is calculated by comparing unigrams from the generated reports with the actual ground truth to assess whether the model produces high-quality reports. SBERT is used to measure the semantic similarity between the model-generated reports and the ground truth, showing how closely the generated content matches the actual reports.

\subsection{Answering RQ1: Fine-tuned models vs. ChatGPT 4o}

As shown in the Figure \ref{fig:RQ1nRQ2} the fine-tuned Qwen2.5 model demonstrates superior performance across all metrics compared to other fine-tuned models. Specifically, the fine-tuned Qwen model achieved a CTQRS score of 77\%, marking a significant improvement of 14\% over its llama model's score of 63\% and improvement of 5\% as compared to ChatGPT. This enhancement trend is consistent across SBERT and ROUGE-1 scores, where the fine-tuned Qwen model outperforms all other fine-tuned models. 

A key factor contributing to this improvement is Qwen's implementation of the Grouped-Query Attention (GQA) mechanism. This advanced attention mechanism provides a notable improvement over the standard attention mechanisms employed by similarly sized models, such as Mistral.
 Additionally, the fine-tuned Qwen model demonstrates a remarkable capability in generating high-quality bug reports from unstructured reports, surpassing lower-parameter models, performing close to SOTA model ChatGPT in post-fine-tuning performance
\cite{qwen2025qwen25technicalreport}.\\ \\
\noindent\fcolorbox{black}{lightgray}{%
    \parbox{.98\linewidth}{%
        \textit{RQ1:} Comparing \textit{Fine Tune and ChatGPT} shows fine-tuned models Qwen and Mistral models comparable to ChatGPT, achieving CTQRS scores of 77\% and 71\%, respectively, compared to 75\% for ChatGPT. Additionally, both fine-tuned models surpass ChatGPT in ROUGE Score, with Qwen and Mistral attaining scores of 0.64 and 0.62, respectively, against 0.44 for ChatGPT.

    }%
} 
\vspace{-3mm}
\subsection{Answering RQ2: Generalizability of Fine-tuned models }
We manually curated a dataset of 300 high-quality reports from the publicly available dataset shared by Song et al. \cite{BEE}. These reports were processed through our pipeline, following the methodology outlined in Section \ref{Methodology}, and evaluated using the same approach as RQ1.

Our results, presented in Figure \ref{fig:RQ1nRQ2}, indicate that fine-tuned Qwen performed comparably to the ChatGPT model achieving 70\% CTQRS score while ChatGPT achieved 73\%. These findings highlight the effectiveness of fine-tuning for specific tasks, demonstrating that task-specific fine-tuned models can achieve similar performance to state-of-the-art (SOTA) models at a lower compute cost.

Additionally, we observed that ChatGPT’s verbose text generation negatively impacted the overall evaluation score. Furthermore, three-shot prompting with ChatGPT outperformed zero-shot prompting, suggesting that providing more examples in the prompt improves evaluation scores.
Further supporting this trend, Pham et al. \cite{Pham2023} emphasized that while ChatGPT can be expensive to deploy for specific natural language generation tasks, fine-tuning smaller models on high-quality, in-domain datasets can lead to superior performance.\\ \\
\noindent\fcolorbox{black}{lightgray}{%
    \parbox{.98\linewidth}{%
        \textit{RQ2: }Comparing \textit{fine-tuned models on other OSS projects}, reveals that the fine-tuned Qwen model achieved a robust 70\% CTQRS score, followed by Mistral with 64\%. Significantly, this outperforms Llama3.2's score of 55\% in CTQRS. These outcome emphasizes that the performance benefits observed with fine-tuned models are not limited to a single dataset. Instead, it indicates a valuable degree of generalizability, suggesting that fine-tuning provides a broadly effective strategy for enhancing model performance across diverse datasets.
    }%
}  
\vspace{-1mm}
   \subsection{Answering RQ3: Mapping and Missing information Detection}

To determine the missing information score, we systematically masked different sections of the unstructured reports in the test dataset, including Steps to Reproduce, Actual Behavior, and Expected Behavior. This approach allowed us to evaluate whether the model could accurately identify if the report miss any information.

As shown in Figure \ref{fig:RQ3}, the model struggled to detect Actual Behavior and Expected Behavior in approximately 45–50\% of cases. Instead, it inferred the missing details based on the available context.However, the Steps to Reproduce section was correctly identified in over 70\% of cases for the Qwen and Mistral models.

To evaluate the Mapping Score, we compared the JSON output of the model for each section against the corresponding section in the actual report. As illustrated in Figure \ref{fig:RQ3}, Actual Behavior and Expected Behavior and Actual Behavior were mapped more accurately, achieving a ROUGE score of 0.72 for the Qwen 2.5 model. This higher scores is likely due to the shorter length of these sections. In contrast, the Steps to Reproduce section, being more detailed and lengthy, had a lower ROUGE and METEOR score of 0.52 \& 0.49. \\ \\ 
\noindent\fcolorbox{black}{lightgray}{%
    \parbox{.98\linewidth}{%
        \textit{RQ3:} The fine-tuned Llama3.2 model demonstrated a slight improvement over other fine-tuned models in identifying missing information, particularly with respect to Actual Results and Expected Results. Our manual analysis revealed that the Qwen and Mistral models tend to generate missing information by inferring from the available context rather than explicitly flagging it as missing to the user. We saw model-frequently generated content for Actual Behavior opposite of the original information provided in Expected Behavior and vice-versa. However, it accurately highlighted when the 'Steps to Reproduce' section was missing for the majority of the samples.

The mapping of unstructured reports to structured formats was performed efficiently by the Qwen and Mistral models. However, we observed a decline in performance, with ROUGE scores dropping from 0.72 for mapping Actual Behavior to 0.52 for Steps to Reproduce. Manual analysis indicated that this decrease was related to the length and details in the "Steps to Reproduce" section. The models exhibited a tendency to introduce additional information in the Steps to reproduce section, which further contributed to the reduced ROUGE and METEOR scores.   
  
    }%
}

\begin{figure*}
    \centering
    \includegraphics[scale=.45]{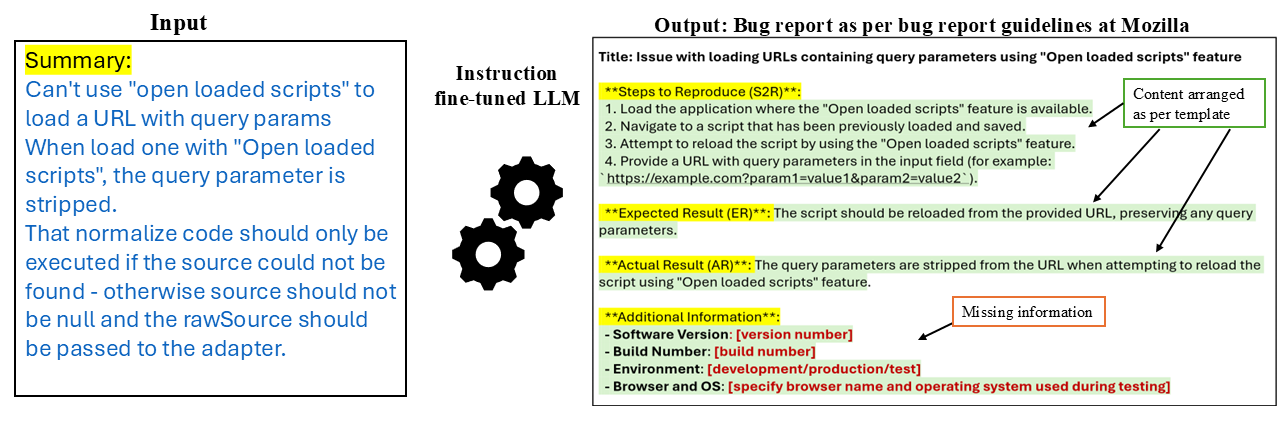}
    \caption{Sample output from our proposed solution approach: Content gets automatically organized as per the template, and missing information is also displayed for further editing for bug reporter}
    \label{fig:solutionEx}
    \vspace{-3mm}
\end{figure*}
\section{Discussion}
\label{Dis}

Our analysis demonstrates that utilizing small language models can achieve performance comparable to state-of-the-art (SOTA) models such as ChatGPT. In addition to their competitive performance, these open-source models offer several advantages, including reduced computational requirements, enhanced scalability, and improved data privacy by mitigating concerns related to proprietary data usage.  

In our study, we have used on the unstructured reports generated (pseudo-ground truth data) using Llama3 model. Llama3 model because of its exceptional performance, open-source nature, transparency, and scalability, which allow us to access, modify, and understand the underlying model architecture and training processes, making it popular among researchers. Due to the lack or absence of a dataset large enough for instruction fine-tuning LLMs, Llama3 provides a valuable alternative for our study. 

% However, we tested our model using the dataset from \cite{multi_sum}, a manually curated 36 bug reports and summary pairs from four open-source projects: Eclipse, Mozilla, KDE, and Gnome.
Our approach demonstrated that LLMs can adapt to various projects and repositories without extensive retraining. This makes our fine-tuned model a valuable tool for different open-source projects, enhancing collaboration and efficiency in software development. The ability of LLMs to understand and generate natural language makes them particularly suited for tasks like bug report generation, where clear and precise communication is essential.

% Traditional methods often require manual effort to write and review bug reports, which can be time-consuming and prone to errors for both developers and reporters. By using an LLM, we can significantly reduce the time and effort involved while maintaining consistency and accuracy in the reports.

Our LLM models, fine-tuned on the pseudo-ground truth dataset successfully generated bug reports in the required format for corresponding unstructured report. Additionally, the model successfully identified and highlighted any missing information according to the bug report template. One such example is shown in Figure \ref{fig:solutionEx}. This ability to generate and evaluate bug reports highlights the utility of LLMs in streamlining the bug-reporting process. The novelty of our approach lies in leveraging LLMs to create bug reports that adhere to bug report templates automatically and ensure they meet specific formatting and informational standards.

% \vspace{-10pt}
% \vspace{-1em}
\vskip -10px
\section{Threats to validity} 
\label{TTV}

In this section, we list various limitations of our study and explain how we address them. \\
\textbf{Internal threats:} 
Our training dataset consists of 3,162 bug reports, which might raise concerns related to sample adequacy for analysis and model fine-tuning. However, Majdik et al. \cite{Majdik2024} demonstrated that a training set of around 2,500 samples can significantly enhance performance in domain-specific tasks like named entity recognition, summarization, and text generation.

{Construct threats:} The eports could introduce biases or inconsistencies in the model's outputs. If the prompts are not carefully crafted to align with the desired output format or guide the model to generate unexpected responses, this could skew the results. 
 Large Language Models like Llama 3 can sometimes hallucinate or produce incorrect unstructured reports, which may impact the accuracy of bug reports and influence our results. Additionally, these models are highly dependent on the prompts given to them; slight changes in the input can lead to significantly different outputs. This sensitivity introduces variability and potential bias into the data generation process, affecting the reliability of our conclusions. We mitigate this threat by carefully designing our prompts using the Alpaca- LoRA template \cite{Tloen} and keeping them consistent throughout the fine-tuning process. \\
 Furthermore, LLMs are prone to hallucination, generating plausible-sounding but factually incorrect or unsupported details—especially when inferring missing sections. In our manual review, we observed instances where the model filled in fields (e.g., actual vs. expected behavior) with contextually plausible yet inaccurate content. Future work will incorporate human validation steps and automatic consistency checks to detect and filter out hallucinated information.
To mitigate the risk of data leakage, we mined the dataset of recent bug reports from bugzilla from November 2024, ensuring that the LLMs had not been exposed to similar data during their pre-training.\\
While computing evaluation measures, external variables such as differences in the length of the bug report generated can lead to unfair comparisons in n-gram overlap metrics, as models producing lengthier or detailed reports may provide helpful information that does not align with well-structured bug reports, thereby lowering their scores. We tried to mitigate this issue by introducing additional grading metrics i.e SBERT and CTQRS.  Furthermore, the quality of data preparation and reference unstructured reports (pseudo-ground truth data) has an impact on the overall results.

\textbf{External threats:} The models, fine-tuned specifically for bug reports belonging to the Mozilla family of projects and bug report guidelines, thus, may not perform as effectively or as relevant in other organizational contexts or with different types of reports, limiting their broader applicability until they are fine-tuned per the organization's guidelines. However, generic testing on samples sourced from \cite{multi_sum} showed encouraging results, thus mitigating this threat to a large extent.

\textbf{Construct threats:} The evaluation metrics ROUGE-1, CTQRS and METEOR emphasize structural or lexical overlap and can underestimate semantic quality for example, penalizing a perfectly clear paraphrase that uses different wording. To address this, we plan to integrate human‑judged evaluations and explore specialized metrics that better capture developer‑perceived clarity, usefulness, and actionability of bug reports in future work.
Incorporating human evaluations or developing specialized metrics tailored to bug report generation could offer a more accurate assessment, which will be part of our future work.

\vspace*{-8pt} 
\section{Related work} \label{reWork}
We discuss related work with respect bug report quality and instruction fine-tuning of LLMs as follows. 

\textbf{Bug Report Management (BRM)}:\label{BRM} Bug report management has evolved through studies focusing on quality assessment and automation of software engineering tasks. Bettenburg et al. \cite{good_report} were the first who explore this issue and analyzed the summary, steps to duplicate, and test cases in the bug report using traditional ML techniques to focus on the quality aspects of bug reports. Zanetti et al. \cite{zanetti2013categorizing} further checked for duplicate images to ensure the quality of the description text and developed methods to assess report validity. Chaparro et al. \cite{steps_to_repo} explored  quality assessment of S2Rs using grammar parsing, neural sequence labelling, and matching of S2Rs to graph-based execution models of the Android application

Since the advancement in NLP, bug report Summarization has been explored widely. Kou et al. \cite{bGS} used three sentence significance factors, i.e. believability, sentence-to-sentence cohesion, and topic association, to summarize the bug report. Later, Xiang and Shao \cite{summlla} proposed SumLLaMA, which used LLMs to generate summaries from bug reports.
% Furthermore, Ma et al. \cite{title_gen} introduced AttSum and explored title generation for bug reports based on its summary.  

The significant manual effort required for bug triage and resolution \cite{guo2010characterizing,good_report}.
Breu et al. \cite{breu2010information}  explored the information requirements in bug reports by studying the questions posed in 600 bug reports from the Mozilla and Eclipse projects.
Additionally,  incomplete or unclear S2Rs pose a big challenge for automated methods to generate test cases from bug reports \cite{fazzini2018automatically,karagoz2017reproducing}. While multiple NLP approaches have been explored, as discussed above, the use of instruction fine-tuning on large language models (LLMs) to improve bug report quality has not yet been investigated. This represents a significant research gap, which our study aims to address.

\textbf{Bug report quality}: It has been widely studied, with various methods proposed to evaluate and improve it \cite{good_report,hao2019ctras}. Many of these methods use heuristic rules and expert insights to identify key details in bug reports. He et al. \cite{he2020deep} introduced a convolutional neural network (CNN)-based approach to classify bug reports as valid or invalid using only textual data, such as summaries and descriptions. Similarly, Chen et al. \cite{chen2018automated} leveraged natural language processing and quantifiable indicators to assess bug report quality automatically. Building on this, we adopted the \textbf{CTQRS} framework by Zhang et al. \cite{CTQRS}, which is a bug-report quality assessment framework that systematically scores bug reports by combining morphological, relational, and analytical indicators through dependency parsing. Morphological indicators (size, readability, punctuation) capture structural and linguistic aspects; relational indicators (itemization, complete environment info, screenshots) examine whether each standard field is properly provided; and analytical indicators (interface elements, user behavior, system defects) tap into deeper semantics by checking how clearly a report describes UI elements, actions, and defect details. Each indicator is linked to desirable properties of atomicity, completeness, conciseness, understandability, and reproducibility and is assigned rule-based weights to compute an overall quality score with a maximum score of 17 points.

\textbf{Instruction fine-tuning:}\label{IT}
Large pre-trained language models are capable of executing a diverse variety of generative tasks utilizing human-annotated instruction data. Notable examples of such tasks include story writing \cite{yuan2022wordcraft}, email drafting, the design of menu systems \cite{kargaran2023menucraft}, poetry composition \cite{chakrabarty2022help}, code generation \cite{muennighoff2023octopack}, and the creation of food recipes \cite{h2020recipegpt}. Taking inspiration from these studies, we explore instruction fine-tuning.

\section{Conclusion and Future Work} \label{CnF}

% Our experiments showed that Qwen 2.5 consistently outperformed other fine-tuned models (Mistral and Llama 3.2) and achieved close to ChatGPT in terms of bug report quality scores (CTQRS), ROUGE, and SBERT metrics. Moreover, the approach generalized effectively to bug reports from different projects e.g. Eclipse, GCC, and Apache, indicating its potential to streamline software maintenance across diverse domains. 
In this paper, we demonstrated how instruction fine-tuned LLMs can automatically convert casual, unstructured bug reports into well-structured ones that closely follow standard templates. Our experiments revealed that Qwen 2.5, when fine-tuned, consistently outperformed other open-source models (Mistral 7B and Llama 3.2) across multiple evaluation metrics, achieving a CTQRS score of 77\% , ROUGE-1 score of 0.64, and SBERT similarity of 0.82. These results underscore its ability to generate structured reports that closely align with human-written outputs, matching the performance of proprietary models like ChatGPT (75\% CTQRS in 3-shot learning). 

 A key contribution of this work lies in addressing the challenge of inconsistent bug report quality, which hinders software maintenance efficiency. By leveraging instruction fine-tuning, our approach ensures that critical components such as Steps to Reproduce, Expected \& Actual Behavior are systematically captured, even when unstructured input lack explicit details. This capability reduces developer effort by clarifying ambiguous reports and helps reporters identify and supply missing information.

Furthermore, our cross-project evaluation highlighted the model’s robust generalization. When tested on bug reports from diverse ecosystems (e.g., Eclipse, GCC, Apache), Qwen 2.5 maintained a CTQRS score of 70\% , proving its adaptability to varying project contexts. Thus showcasing the efficacy of instruction fine-tuning on open-source models, we provide evidence that such models can serve as cost-effective, privacy-preserving, scalable alternatives to proprietary systems like ChatGPT, without compromising much on performance.

As future work, we plan to enhance the model’s capabilities by integrating richer data sources (e.g., error snapshots, logs, and code snippets), exploring advanced fine-tuning techniques such as QLoRA, expanding support for additional bug-tracking platforms (e.g., GitHub and Jira), and developing a real-time tool that proactively assists reporters in providing missing details. These improvements aim to enhance the bug-reporting experience and minimize developers' manual effort.

% In this research, we demonstrated how instruction fine-tuned LLM models can be leveraged to automatically transform unstructured bug reports into well-structured bug reports. Additionally, we showed that the LLM models trained on pseudo-ground-truth data performed well for smaller project data from the Bugzilla repository and also on another bug report and summary dataset from Eclipse and KDE, emphasizing the generalizability of our approach. Since the quality of bug reports plays a pivotal role in the software development lifecycle, significantly impacting the efficiency of bug resolution processes, mainly bug triage, bug assignment and the overall quality of software products, we believe we address the first step of solving a bug which is ensuring well-structured bug report filing. 

\bibliographystyle{ACM-Reference-Format}
\bibliography{sample-base}

\end{document}